\documentclass[prl,twocolumn,amsmath,amssymb,10pt,aps,longbibliography,superscriptaddress,citeautoscript,bibnotes,footinbib]{revtex4-1}
\usepackage{feynmp}
\usepackage{amsmath}
\usepackage{graphicx}
\usepackage{rotating}
\usepackage{multirow}
\usepackage{latexsym}
\usepackage{textcomp}
\usepackage{verbatim}
\usepackage{color}
\usepackage{bm}
\usepackage{subfigure}
\usepackage{everysel}
\usepackage{keyval}
\usepackage{ragged2e}
\usepackage{dsfont}
\usepackage{amssymb}
\usepackage{enumitem}
\usepackage{pstricks}
\usepackage{mathtools}
\usepackage{braket}
\usepackage{slashed} 
\usepackage[colorlinks=true]{hyperref}

\newcommand{\osum}{{%
    \setbox0\hbox{\circ}%
    \rlap{\hbox to \wd0{\hss\sum\hss}}\box0
}}

\begin{document}

\title{Symmetric-Gapped Surface States of Fractional Topological Insulators}   % type title between braces

\author{Gil Young Cho}         % type author(s) between braces
\affiliation{School of Physics, Korea Institute for Advanced Study, Seoul 02455, Korea}
\affiliation{Department of Physics, Korea Advanced Institute of Science and Technology, Daejeon 305-701, Korea}

\author{Jeffrey C. Y. Teo}
\affiliation{Department of Physics, University of Virginia, Charlottesville, VA 22904 USA}

\author{Eduardo Fradkin}
\affiliation{Department of Physics and Institute for Condensed Matter Theory, University of Illinois, 1110 W. Green St., Urbana, Illinois 61801-3080, USA}
\date{\today}    % type date between braces
 
\begin{abstract} 
We construct the symmetric-gapped surface states of a fractional topological insulator with electromagnetic $\theta$-angle $\theta_{em} = \frac{\pi}{3}$ and a discrete $\mathbb{Z}_3$ gauge field. They are the proper generalizations of the T-pfaffian state and pfaffian/anti-semion state and feature an extended periodicity compared with their  of ``integer" topological band insulators counterparts. We demonstrate that the surface states have the correct anomalies associated with  time-reversal symmetry and charge conservation.  
\end{abstract}

\maketitle
\textbf{Introduction:} The three-dimensional topological band insulator\cite{Rev1, Rev2, Rev3, RMP} is an electronic topological phase. Its discovery embodies  the remarkable progresses in our understanding of the interplay between symmetry and topology of quantum states of matter. Its topological nature manifests spectacularly as a single gapless Dirac fermion living at its boundary, which is otherwise impossible to exist. It has been thought for some time that a single gapless Dirac fermion is the only allowed surface state respecting  time-reversal and charge conservation symmetries. However, surprisingly, it turns out that there is another option \cite{VS}: the surface can be gapped while respecting the symmetries, at the cost of introducing  topological order, resulting in the T-pfaffian state and the pfaffian/anti-semion state \cite{GS1,GS2,GS3,GS4}.

In this paper, we will consider the surface of  {\em  a fractional topological insulator} (FTI). The fractional topological insulator\cite{FTI1,FTI2,FTI3,FTI4,FTI5,FTI6} is a symmetry-enriched topologically ordered state of matter in three spatial dimensions, which supports anomalous surface states protected by time-reversal symmetry and charge conservation. The simplest 3D FTI \cite{FTI3} contains fractional excitations such as gapped charge-$\frac{1}{3}$ fermions and $\mathbb{Z}_3$ gauge fluxes. It is characterized by a term in the effective action for the electromagnetic response of the {\em bulk} with a fractional axion angle $\theta_{em} = \frac{\pi}{3}$ \cite{FTI3} 
\begin{align}
\mathcal{L}_\theta = \frac{\theta_{em}}{32\pi^2} \varepsilon^{\mu\nu\lambda\rho}F_{\mu\nu}F_{\lambda \rho} = \frac{1}{96\pi} \varepsilon^{\mu\nu\lambda\rho}F_{\mu\nu}F_{\lambda \rho}.
\label{Axion}
\end{align} 
This state can be constructed by fractionalizing the electron into the three charge-$\frac{1}{3}$ fermionic partons, i.e., $\Psi_e = \psi_1 \psi_2 \psi_3$,  which, at the mean field theory level, is described by the topological band insulator. The fractionalization of the physical electron into the multiple fermionic partons introduces unphysical states in the Hilbert space and those states need to be projected out. This projection is done efficiently by introducing a $\mathbb{Z}_3$ gauge field and make the partons $\psi_j$ carry the unit charge under this gauge field, i.e., under the gauge transformation, the parton transforms as $\mathbb{Z}_3 : \psi_j \to \omega \psi_j$ with $\omega^3 =1$. On the other hand, the electron is locally gauge-invariant, i.e.,  $\mathbb{Z}_3 : \Psi_e \to \Psi_e$, as it should be. Here we will asume that the $\mathbb{Z}_3$ gauge theory is realized in its deconfined phases \cite{comment1,fradkin-shenker-1979}. Several works on theoretical constructions of 3D FTIs have been written, and some of the  physics of the bulk states is by now reasonably well understood.
%However, concrete physical microscopic  models of these states are still lacking.} 

Compared to the bulk, the surface states of FTIs have been less studied and are not well understood, largely because of the strong interactions required for these states to occur. The surface of a FTI  is intrinsically strongly-correlated and thus the fate of the surface Dirac fermions, which result in the mean-field description, is not \textit{a priori} clear. In the presence of the strong interactions, there are several scenarios possible to happen. The surface may break the symmetries protecting the gapless-ness spontaneously and be gapped. A more interesting possibility is to have a transition to a phase which is gapped while respecting all the symmetries. This phase is the symmetric-gapped surface state, which lives \textit{only} on this (3+1)-dimensional state with symmetry-enriched topological order. Such a surface state should realize the symmetries in an anomalous fashion which cannot be realized within strictly two space dimensions. 

In this paper we construct a gapped state on the surface of a 3D FTI. This state is invariant under the $\mathbb{Z}_3$ gauge symmetry and respects global electric charge conservation and time-reversal invariance. Since it is gapped, this state should be stable against  interactions with moderate strength. This state is the generalization of the T-pfaffian state of the 3D time-reversal-invariant topological insulator \cite{GS3} (see also Ref. \cite{Seiberg-Witten-2016}) to the more general problem of the surface of a 3D FTI. More precisely, we show that the generalization of the symmetric-gapped surface states of the topological insulator have the extended periodicity, which are forced by the $\mathbb{Z}_3$ gauge invariance. This extended periodicity makes the surface of the FTI to have the correct parity anomaly.

The  symmetries of symmetric surface states of the 3D FTI, at the quantum level,  are realized anomalously, %be suffer from the correct ``anomalies", 
which implies that this state can only occur  on the surface of a 3D systems this the correct bulk anomaly. The anomaly of the surface that we are mainly concerned in this paper is a fractional parity anomaly with an associate surface Hall conductivity $\sigma_{xy}= \frac{1}{6}$. This anomaly must either  be cancelled by the bulk or by another surface state \cite{Callan-1985,Fradkin-1986,Qi-2008}. For example, the T-pfaffian state\cite{GS3} has the same parity anomaly as the single Dirac fermion, i.e., $\sigma_{xy}=\frac{1}{2}$ \cite{Redlich,Witten}. To see this clearly, we note that the single Dirac fermion alone is not invariant under large gauge transformations, and we need to regularize the theory properly to restore the gauge symmetry at the cost of breaking the time-reversal symmetry, i.e., the properly-regularized theory comes along with a half-level of the Chern-Simons term, $-\frac{1}{8\pi}\varepsilon^{\mu\nu\lambda}A_\mu \partial_\nu A_\lambda$, which explicitly breaks the time-reversal symmetry \cite{Redlich}. 

However, when coupled to the bulk of the topological insulator, time-reversal symmetry at the surface is restored \cite{Witten} by  the axion term in the bulk effective electromagnetic action,
% The bulk of the topological insulator has the so-called axionic electromagnetism 
$
\mathcal{L}_{em} = \frac{1}{32\pi} \varepsilon^{\mu\nu\lambda\rho}F_{\mu\nu}F_{\lambda\rho}, 
$
whose boundary action cancels the half-level Chern-Simons term generated from the regularization of the Dirac fermion. The T-pfaffian state also has the same parity anomaly $\sigma_{xy}=\frac{1}{2}$ which exactly matches this bulk contribution \cite{GS3}. In the fractional topological insulator case, the axion angle Eq.\eqref{Axion} is $\theta_{em} = \frac{\pi}{3}$ implies that the correct boundary state should have a parity anomaly with $\sigma_{xy}=\frac{1}{6}$. Hence, we look for states with $\mathbb{Z}_3$ gauge symmetry, global electric charge conservation, and time-reversal symmetry, and a parity anomaly $\sigma_{xy}=\frac{1}{6}$. 

Here we  construct such symmetric-gapped states with the help of the recently-developed fermionic dualities in (2+1) space-time dimensions \cite{Duality1, Duality2, Duality3, Duality4}. One of the states that we construct is the generalization of T-pfaffian state, that exactly matches the topological order that two of us found previously in an anyon-theoretic construction  \cite{FTI_Surf}.  Various heterostructures of FTI thin films were considered and constrained the possible structures to derive a symmetric-gapped state. Here, we present a field theoretic derivation of this state, and construct other classes of the symmetric-gapped states for the FTI.

%In constructing the surface state, we use the recently-developed (2+1)d duality, which was valuable in finding the T-pfaffian state. Contrast to the original derivations of the T-pfaffian state which use non-trivial anyon condensations, the exotic topological state can be constructed from the duality approach in a straightforward fashion. So far, the duality has been mainly used in clarifying the relationships between the different theories. However, in this paper, we will show that it can be also useful in finding novel states of matter. 

\textbf{Generalization of the T-pfaffian State:} At the level of mean field theory, the surface state of the 3D FTI consists of the three partons, electric charge-$\frac{1}{3}$ Dirac fermions
\begin{align}
\mathcal{L} = &\sum_{j=1}^{3} \bar{\psi}_j i\slashed{D}_{A/3} \psi_j 
\end{align}
where $A$ is the background electromagnetic gauge field. Note that there are no Chern-Simons terms for the $A$ and the $\mathbb{Z}_3$ gauge fields \cite{comment3}.
%One can  show that there are no such terms since contributions  from the bulk axion term Eq.\eqref{Axion} and  from the regularization of the Dirac fermions, to the Chern-Simons-type terms of $A_\mu$ and $\mathbb{Z}_3$ gauge field cancel each other out (see the supplementary materials for details).
As noted above, this theory is incomplete: the partons must also be coupled to a $\mathbb{Z}_3$ {\em dynamical} gauge field to reproduce the correct Hilbert space.
%where each Dirac fermion is coupled to the $\mathbb{Z}_3$ gauge field. 
The fluctuations of $\mathbb{Z}_3$ gauge field are gapped in the deconfined phase and we have suppressed their explicit contribution to the low-energy effective theory. Nevertheless, the requirement of $\mathbb{Z}_3$ gauge invariance will play a key role. On the other hand, two of the three Dirac fermions can become massive without breaking any of the symmetries of the theory and, generically, we are left with only one massless Dirac fermion, whose mean-field effective action is
% because two of the Dirac fermions can be gapped out without breaking the symmetries. Hence, on the surface, after inclusion of the generic symmetry-respecting perturbations, we have}
\begin{align}
\mathcal{L} =  \bar{\psi} i\slashed{D}_{A/3} \psi %\textcolor{blue}{- \frac{1}{24 \pi} \varepsilon^{\mu\nu\lambda}A_\mu \partial_\nu A_\lambda} 
\label{Direct_Surf}
\end{align}
%in the mean-field level. 
We should recall that the Dirac fermion $\psi$ carries the unit  $\mathbb{Z}_3$ charge, even though the gauge field is not shown explicitly in this low-energy theory.

Another surface state can be obtained from the {\em dual theory} of Eq.\eqref{Direct_Surf}. Duality has provided a direct way to gap out the Dirac fermion without breaking symmetries in the topological band insulator case \cite{Duality1, Duality2}, and we will follow the same strategy here. Upon the duality transformation, the dual theory of Eq.\eqref{Direct_Surf} becomes
\begin{align}
\mathcal{L} =  \bar{\chi} i \slashed{D}_{a} \chi + \frac{1}{12\pi} \varepsilon^{\mu\nu\lambda}a_\mu \partial_\nu A_\lambda %\textcolor{blue}{- \frac{1}{24 \pi} \varepsilon^{\mu\nu\lambda}A_\mu \partial_\nu A_\lambda}
\label{EffectiveDirac}
\end{align}  
This duality is a short-hand representation which is sufficient for present purposes \cite{comment2}.
%A more correct version of the duality \cite{Duality2} can be used and yields the same answer (see the supplementary materials).} 
Here we would like to introduce the gap while preserving the symmetries. We first introduce an s-wave pairing field, i.e., a singlet pairing in the spinor index, to $\chi$ fermions. Note that here the pairing is dynamical and originates from the strong correlations, contrary to the proximity effect in the Fu and Kane model \cite{Fu-2008}. Because the $\chi$ fermion is explicitly electrically charge neutral, the s-wave paired state of Eq.\eqref{EffectiveDirac} respects  time-reversal symmetry and  charge conservation. This is the T-pfaffian state of the parton $\psi$.

We review a few facts about the T-pfaffian state, needed for our construction. The effective low energy theory of the  T-pfaffian has a charge sector and a neutral (Ising) sector \cite{GS3}. The excitations of the charge sector are labelled by their vorticity $k$ mod 8, and are charge-$\frac{k}{4}$ anyon excitations of the filling $\nu=\frac{1}{8}$ state of the charge-$2$ boson. The excitations of the Ising sector are the abelian boson $I$, fermion $f$, and the non-abelian anyon $\sigma$. In the T-pfaffian state, excitations with  even vorticity, $k= 2n$, are bound with the abelian anyons $I$ and $f$ of the Ising sector, and  excitations with  odd vorticity $k=2n+1$ are bound to the non-abelian anyon $\sigma$. The resulting states are respectively denoted below as $I_k$, $f_k$ and $\sigma_k$. $I_8$ is a  boson braiding trivially with all the other anyons. Its presence truncates the spectrum of the theory to $12$ excitations \cite{GS3, Duality1}. The main difference from the T-pfaffian state of the topological band insulator is in the electric charge carried by the excitations: The vorticity $k$ excitation carries the electric charge $\frac{k}{12}$ instead of $\frac{k}{4}$ as in the charge of excitations of the topological band insulator. 

%Details of the anyon content and symmetry charges of the anyons are reviewed in the supplementary material. 
%\begin{table}
%\begin{tabular} {c|c|c|c|c|c|c|c|c}
%\hline
%k & 0& 1 & 2 & 3 & 4 & 5 & 6 & 7 \\ \hline \hline
%$Q_\text{em}$ & 0& $\frac{1}{12}$ & $\frac{1}{6}$ & $\frac{1}{4}$  & $\frac{1}{3}$  & $\frac{5}{12}$ &  $\frac{1}{2}$ &  $\frac{7}{12}$  \\ \hline
%$\mathbb{Z}_3$  & 0 & 1 & 2 & 0& 1 & 2 & 0 & 1  \\ \hline \hline
%$I_{k}$ & 1&  {} & $\times$ &  {}  & $-1$  & {}  &  $\times$ &   {} \\ \hline
%$f_{k}$ & 1&  {} & $\times$ &  {}  & $-1$  & {}  &  $\times$ &   {} \\ \hline
%$\sigma_k$ & {} &  1 & {} &  -1  & {}  & -1  &  {} &   1 \\ \hline
%\end{tabular}
%\caption{Anyon content and gauge charges of the T-pfaffian state of Eq.\eqref{Direct_Surf}. From the third row to the fifth row, we have written out the signs of $\mathcal{T}^2$ acting on the excitations. For $k=2$ and $k=6$, $I_k$ and $f_k$ are exchanged under $\mathcal{T}$ as in the T-pfaffian. Here $\{I, \sigma, f\}$ are from the Ising sector. The topological spins of the excitations are identical to the T-pfaffian state.}
%\label{TpfaffianCharge}
%\end{table}

Without the $\mathbb{Z}_3$ gauge field, this T-pfaffian state of the parton could have been a legitimate symmetric surface state. However, here we need to be more careful because of the internal $\mathbb{Z}_3$ gauge field. To see this, we first identify the $\mathbb{Z}_3$ gauge charge of the excitations. We first assign the $\mathbb{Z}_3$ gauge charge $q$ to the smallest excitation, $\sigma_1$. Then, the excitation of the vorticity $k$ carries the $\mathbb{Z}_3$ gauge charge $k \times q$ mod 3. Since the fermion $f_{4}$ has the same quantum numbers as the parton $\psi$ \cite{GS3, Duality1}, i.e., electric charge-$\frac{1}{3}$ and unit $\mathbb{Z}_3$ charge, we obtain the constraint 
\begin{align}
4q = 1 \text{ mod } 3, 
\label{Charge}
\end{align}
One solution to Eq.\eqref{Charge} is  $q=1$. (We will come back below to the other solution $q = \frac{1}{4}$ mod 3.) From this, we read how the excitation $V_k$ of the vorticity $k$ transforms under the $\mathbb{Z}_3$ gauge transform, i.e., $\mathbb{Z}_3 : V_k  \to \omega^{k} V_k$.

This has a striking effect on the anyon theory: \textit{the T-pfaffian of the parton $\psi$ breaks $\mathbb{Z}_3$ gauge symmetry} because the supposedly-`transparent' boson $I_8$ \cite{GS3} transforms non-trivially under the $\mathbb{Z}_3$ gauge transformations, i.e., $\mathbb{Z}_3 : I_8 \to \omega^2 I_8$.  The boson $I_8$ is non-local because it has a non-trivial braiding phase with the $\mathbb{Z}_3$ flux. Hence, the anyon contents {\em can no longer} have period $8$ if the internal gauge invariance $\mathbb{Z}_3$ is to be respected. If we enforce the periodicity to be $8$, then we need to break $\mathbb{Z}_3$ gauge symmetry completely and remove the $\mathbb{Z}_3$ flux from the excitation spectrum. Physically, the boson $I_8$ corresponds to the pair field of the fermion $\psi$, which carries charge-$2$ under $\mathbb{Z}_3$ gauge group and electric charge $\frac{2}{3}$. Thus, the T-pfaffian of the parton is not compatible with the internal $\mathbb{Z}_3$ gauge symmetry.

%which, in the T-pfaffian state, is identical  to the vacuum $I_0$,

There are two options to restore the $\mathbb{Z}_3$ gauge symmetry to this state. One   is  simply to remove the pairing in the $\chi$ fermions and to go back to the metallic state of Eq.\eqref{Direct_Surf}. The other option is to enter into a new topological state, and this is the direction that we pursue. The new topological state features an extended periodicity of the anyon contents enforced by $\mathbb{Z}_3$ gauge symmetry.

We start with noting that the triple of $I_8$, i.e., $I_{24} \sim (I_8)^3$, is neutral under the $\mathbb{Z}_3$ gauge field, and, thus, it has trivial braiding phases with all the anyons, including the $\mathbb{Z}_3$ gauge fluxes. Thus, we can truncate the anyon contents at $k=24$ instead of at $k=8$. Therefore,  the $\mathbb{Z}_3$ gauge symmetry can be restored  simply by extending the periodicity of the anyon content of the vorticity from $8$ to $24$. For this state,  time-reversal symmetry as well as  charge conservation are inherited directly from the ``parent" T-pfaffian state of the parton $\psi$. Hence, this state respects all the required symmetries to be the legitimate surface state of the fractional topological insulator. 

%\textcolor{red}{It is easy to see that giving up the $\mathbb{Z}_3$ gauge symmetry, the anyon content reduces to the T-pfaffian state of the parton $\psi$.}

We now investigate the consequences of the extended periodicity, $k \sim k+ 24$. In this theory, the topological spins and the action of  time-reversal symmetry $\mathcal{T}$ still repeat with period $8$. The charges are assigned as follows
\begin{align}
Q_{em, k} = \frac{k}{12}, ~\text{and} ~\mathbb{Z}_3 : V_k  \to \omega^{k} V_k, 
\end{align}
where $V_k$ represents the anyons with vorticity  $k$, with $k \sim k+24$. Also,  $I_k$ and $f_k$ with $k\equiv2$ mod 4 are exchanged under time reversal. For example, there are two types of anyons $V_k$, i.e., $I_{18}$ and $f_{18}$, carrying  electric charge $\frac{3}{2}$, and  are exchanged under the time-reversal symmetry, as in the usual T-pfaffian state. There are two excitations to which we pay a particular attention. The first is  the electron quasiparticle $\Psi$, i.e., $f_{12}$, which carries  electric charge $1$, is neutral under $\mathbb{Z}_3$, and has $\mathcal{T}^2 = -1$. The second is the (singlet) Cooper pair of  electrons, which is identified with $I_{24}$: a boson that has  electric charge $2$ and is a Kramers singlet.

We now come to the parity anomaly. Without referring back to the field theoretic derivation, we can read off the anomaly directly from the anyon content of the theory. This way of reading off the anomaly will be useful when discussing the generalization of the pfaffian/anti-semion state. In the case of the T-pfaffian state of the topological band insulator, the period is $8$ and the vacuum is identified with the charge-$2$ boson $I_8$. This implies that $\sigma_{xy} = \nu \times Q^2 = \frac{1}{8} \times 2^2 = \frac{1}{2}$, where $\nu$ is the inverse of the periodicity (more precisely, it is the ``filling" of the bosonic charge sector, or the level of the charge-sector Chern-Simons term) and $Q$ is the charge of the transparent boson. Hence, we see that the T-pfaffian state has the correct parity anomaly $\sigma_{xy}=\frac{1}{2}$. Now, for the surface state of the {\em fractional topological insulator}, the period $24$ with  charge-$2$ boson $I_{24}$ implies that the surface has the charge response with Hall conductivity $\sigma_{xy} = \nu \times Q^2 = \frac{1}{24} \times 2^2 = \frac{1}{6}$. This is precisely the expected parity anomaly of the FTI! A more accurate statement is that the charge sector of this anyon theory is  $U(1)_{24}$, as was  shown explicitly in Ref. \cite{FTI_Surf}. 

It is now clear that the other solution $q=\frac{1}{4}$ to Eq.\eqref{Charge} generates the same anyon content as the  $q=1$ solution because, in the above analysis, only the $\mathbb{Z}_3$ gauge charge of  $I_8$ is important. Obviously, the $\mathbb{Z}_3$ gauge charge of $I_8$ is the same in both the solutions. The fractionalization of $\mathbb{Z}_3$ gauge charge by $q=\frac{1}{4}$, which  extends the $\mathbb{Z}_3$ gauge theory to the $\mathbb{Z}_{12}$ gauge theory \textit{only at the surface}, does not break the $\mathbb{Z}_3$ gauge symmetry, charge conservation, and time-reversal symmetry. Hence, this state is also another legitimate surface state of the fractional topological insulators. The two solutions, $q= \frac{1}{4}$ mod $3$ and $q= 1$ mod $3$, can be distinguished by the braiding with the $\mathbb{Z}_3$ flux and the surface excitations because the surface excitations carry the $\mathbb{Z}_3$ gauge charge and thus have non-trivial statistics with the flux.  

In our system, the $\mathbb{Z}_3$ gauge theory is not coupled with the electromagnetic field. For instance, a $\mathbb{Z}_3$ flux does not necessarily carry a non-trivial magnetic flux. Furthermore, because of the time-reversal symmetry, when it intersects the symmetric surface, it does not carry an electric or gauge charge. 

We now discuss the topological degeneracy on the open 3-manifold $D^2\times S^1$ of this fractional topological insulator with the symmetric-gapped topologically ordered surface, where $D^2\times S^1$ is the \textit{filled} spatial torus. We note that there are only one non-contractible loop along $S^{1}$ along which we have three possible degeneracies labelled by the $\mathbb{Z}_3$-charge Wilson loop around this $S^{1}$. On the other hand, given a $\mathbb{Z}_3$ charge, there are six possible anyonic loops living purely on the surface of $D^2\times S^1$. So, the total  degeneracy is $18$.

%\textbf{I do not understand these two short paragraphs. Also the preceding paragraph needs refinement: I assume that the ``anyonic self-statistical angle" is just the Dehn twist of the anyon which is complex in  a chiral theory, isn't it?}

\textbf{Relation with the paired FQH state at filling $\nu=\frac{1}{6}$:} It has been conjectured from the link between the half-filled composite Fermi liquid and the surface of the topological band insulator \cite{Son_CFL,Wang_Senthil_CFL}  that the particle-hole symmetric version of the T-pfaffian state, the PH-pfaffian \cite{Son_CFL}, can be realized in a half-filled Landau level. This state is essentially equivalent to the T-pfaffian in terms of  symmetries and excitations but  time-reversal symmetry is replaced by the particle-hole symmetry of  the half-filled Landau level (in the large cyclotron energy limit).
We can ask if our exotic surface state of the FTI can be realized in a Landau level. From the charge response $\sigma_{xy}= \frac{1}{6}$ of the surface state, it is natural to compare this state with a putative paired FQH state at $\nu=\frac{1}{6}$. However, contrary to the half-filled case, we do not have an obvious  particle-hole symmetry at  $\nu=\frac{1}{6}$. This implies that the surface state of the FTI does not have a natural partner in a fractionally-filled Landau level. 
%This is sharply different from the half-filled case, where the particle-hole symmetry is emergent when the cyclotron gap becomes large enough.
In fact, the excitations of the paired FQH state at the filling $\nu=\frac{1}{6}$ can be generated by tensoring of the charge sector $U(1)_{24}$ and the neutral Ising sector, quotiented by an extended symmetry. However, as discussed in the work \cite{FTI_Surf}, to restore the time-reversal symmetry   another neutral $\mathbb{Z}_3$ sector is needed, which is absent in the paired FQH state. 

\textbf{Other Symmetric-Gapped Surface States:} On establishing the generalization of the T-pfaffian state at the surface of FTIs, we now address if we can construct the generalization of another symmetric-gapped state of topological band insulator, i.e., a pfaffian/anti-semion state.\cite{GS1,GS3} 
For this, we note that the essential step in our generalization of the T-pfaffian state is to identify $f_{4}$, the electron in the topological band insulator case, with the minimal parton $\psi$ carrying unit $\mathbb{Z}_3$ gauge charge. This state breaks the internal $\mathbb{Z}_3$ gauge symmetry which can be restored by extending the periodicity of the anyon content from $8$ to  $24$. This strategy, ``extending periodicity" to restore the internal gauge symmetry, straightforwardly generalizes to the other symmetric-gapped surface order, e.g., the pfaffian/anti-semion state. 
This state \cite{GS1,GS3}, which is realized at the surface of the topological band insulator, respects time-reversal symmetry and charge conservation. Its excitations are labelled by $\{ I_k, I_k s, \sigma_k, \sigma_k s, f_k, f_k s \}$ with  vorticity $k \sim k+8$ (here $s$ is the anti-semion), and carry  electric charge $\frac{k}{4}$. In this state, $f_4$ is the electron, i.e., a Kramers doublet charge-$1$ fermion, and the singlet Cooper pair $I_8$ is ``transparent" to all the anyons, as in the T-pfaffian state. 

On the surface of the fractional topological insulator, we imagine to put the parton $\psi$ first into the pfaffian/anti-semion state. Temporarily ignoring the $\mathbb{Z}_3$ gauge symmetry, we find a theory respecting all the symmetries. The only difference is the electric charge carried by the anyons. Now the excitation with  vorticity $k$ carries  electric charge $\frac{k}{12}$ since the ``elementary" excitation $\psi$ has  fractional electric charge $\frac{1}{3}$. Obviously, on bringing the $\mathbb{Z}_3$ gauge symmetry back into the discussion, we see that  $I_8$, which is supposed to be local in the usual pfaffian/anti-semion state, is no longer local and braids with the $\mathbb{Z}_3$ fluxes since it carries  charge $2$ under the $\mathbb{Z}_3$ gauge field. However, we can restore the $\mathbb{Z}_3$ gauge symmetry by extending the periodicity once again from $8$ to $24$, i.e., $k \sim k +24$. In this state, the charge sector has period $24$ and the identity boson carries  electric charge $2$. Hence the parity anomaly associated with this state is again $\sigma_{xy} = \frac{1}{6}$, the correct anomaly to be on the surface of the fractional topological insulator.  Furthermore, it obeys   time-reversal symmetry and charge conservation,  inherited from the original pfaffian/anti-semion theory.

\textbf{Conclusions and Outlook:} In this paper, with the help of the  fermion-fermion duality we  constructed the symmetric-gapped surface states of  FTIs with  electromagnetic axion angle $\theta_{em} = \frac{\pi}{3}$, whose excitations are the fractional parton and the discrete $\mathbb{Z}_3$ gauge flux. The symmetric-gapped surface states are generalizations of  the T-pfaffian state and the pfaffian/anti-semion state but \textit{with an extended periodicity}. We showed that the surface states respect the required symmetries of charge conservation, time-reversal symmetry, and the $\mathbb{Z}_3$ gauge symmetry, and that they have the correct parity anomaly, i.e., $\sigma_{xy}= \frac{1}{6}$, which matches the axion angle $\theta_{em} = \frac{\pi}{3}$. At the heart of finding these surface states, the identification of the electron in the symmetric-gapped surfaces of topological band insulator by the \textit{non-local} fermionic parton plays an essential role. This requirement forced the extension the periodicity of the anyon content so as to restore the internal $\mathbb{Z}_3$ gauge symmetry.  
Here we focused on the case of FTIs with $\theta_{em} = \frac{\pi}{3}$, but it is straightforward to generalize our construction to the other FTIs with angle  $\theta_{em} = \frac{\pi}{2n+1}$ and a $\mathbb{Z}_{2n+1}$ gauge field. 
We end by noting that  structure we used  (``extending periodicity")   will arise in the construction the symmetric-gapped surface states for other various bosonic and fermionic FTIs. Generically, we expect that they will inherit their global symmetries from their counterparts in the ``integer" bosonic and fermionic topological insulators. 
%We leave the  construction of symmetric-gapped surface states for  broader classes of FTIs, e.g., with continuous gauge groups \cite{FTI1, FTI3}, for a future publication. 

\acknowledgements
We thank F. Burnell, M. Cheng, T. Faulkner, H. Goldman,  M. Metlitski, S. Ryu, N. Seiberg, H. Wang, and P. Ye, for helpful discussions and comments. This work is supported by the Brain Korea 21 PLUS Project of Korea Government (G.Y.C.), Grant No. 2016R1A5A1008184
under NRF of Korea (G.Y.C.), NSF Grant No.~DMR-1653535  at the University of Virginia (J.C.Y.T.), and NSF grant No. DMR 1408713 at the University of Illinois (E.F). G.Y.C. also acknowledges the support from the Korea Institute for Advanced Study (KIAS) grant funded by the Korea government (MSIP).  G.Y.C and J.C.Y.T  thank S. Sahoo and A. Sirota for collaboration in a  previous work,  EF to P. Ye and M. Cheng.  

%\bibliographystyle{apsrev}
%\bibliography{ref_FTI}

%%%%%%%%%%%%%%%%%%%%%%%%%%%%%%%%%%%%%

%%%%%%%%%%%%%%%%%%%%%%%%%%%%%%%%%%%%%%%%%%%%%%%%%%%

\onecolumngrid
\clearpage
\begin{center}
\textbf{\large Supplemental Material for ``Symmetric-Gapped Surface States of Fractional Topological Insulators"}
\end{center}
\begin{center}
{Gil Young Cho, Jeffrey C. Y. Teo, and Eduardo Fradkin}\\
%\emph{Department of Physics, Korea Advanced Institute of Science and Technology, Daejeon 305-701, Korea}
\end{center}
\setcounter{equation}{0}
\setcounter{figure}{0}
\setcounter{table}{0}
\setcounter{page}{1}

\section{Dirac Fermion and Chern-Simons term}
In this supplemental material, we clarify the meaning of the Dirac fermion path integral in comparison with Ref. \cite{Duality2}. In Ref. \cite{Duality2}, when the Dirac fermion action is written down, 
\begin{align}
\mathcal{L} = i \bar{\Psi} \slashed{D}_A \Psi,  
\end{align}
it is understood that this action is well-defined theory by a Pauli-Villars regularization. The contribution from the regularization is hidden in the integration measure of the fermionic field $\Psi$ and does not appear explicitly in the action. The contribution is the $\eta$-invariant term, which is essentially equivalent to the half-level Chern-Simons term, $-\frac{1}{8\pi} \epsilon^{\mu\nu\lambda}A_\mu \partial_\nu A_\lambda$ \cite{Witten,Seiberg-Witten-2016}. Hence, the Dirac fermion action appearing in Ref. \cite{Duality2} is not time-reversal symmetric, or, equivalently, is time-reversal symmetric up to the anomaly $\frac{1}{4\pi} \epsilon^{\mu\nu\lambda} A_\mu \partial_{\nu} A_\lambda$. 

In our convention (which is more familiar with condensed matter community, e.g., this is the convention used in Ref. \cite{Duality1}), when we write the Dirac fermion action,  
\begin{align}
\mathcal{L} = i \bar{\Psi} \slashed{D}_A \Psi,  
\end{align}
\textit{we do not assume a `hidden' contribution from the Pauli-Villars regularization} and this action lives only on the surface of a three-dimensional topological band insulator. Hence, this action is time-reversal symmetric. When regularized properly, this action must be modified as 
\begin{align}
\mathcal{L}_{\text{reg.}} = i \bar{\Psi} \slashed{D}_A \Psi - \frac{1}{8\pi} \epsilon^{\mu\nu\lambda}A_\mu \partial_\nu A_\lambda.
\end{align}
The Chern-Simons term for $A_\mu$ originates from the regularization. To restore the time-reversal symmetry for this regularized Dirac fermion action, we need to attach the bulk of the topological band insulator whose effective action is 
\begin{align}
\mathcal{L}_{\text{bulk}} = \frac{1}{32\pi} \epsilon^{\mu\nu\lambda\rho}F_{\mu\nu} F_{\lambda\rho},  
\end{align} 
which contributes another Chern-Simons term $\frac{1}{8\pi} \epsilon^{\mu\nu\lambda}A_\mu \partial_\nu A_\lambda$ on the surface, which exactly cancels the contribution of the Pauli-Villars regularization fields. 

Here we note that, in the convention of Ref. \cite{Duality2}, the time-reversal symmetric regularized Dirac fermion corresponds to 
\begin{align}
\mathcal{L} = i \bar{\Psi} \slashed{D}_A \Psi + \frac{1}{8\pi} \epsilon^{\mu\nu\lambda}A_\mu \partial_\nu A_\lambda, 
\end{align}
in which only the Chern-Simons term from the bulk is explicitly written out. This bulk term is explicitly cancelled by the regularization contribution on the surface. 

Keeping this in mind, we now show that there is no Chern-Simons term for the background electromagnetic gauge field $A_\mu$ and dynamical $\mathbb{Z}_3$ gauge field in Eq. (2) of the main text. Here we take the $\mathbb{Z}_3$ gauge field to be obtained from the dynamical U(1) gauge field $\alpha_\mu$ by condensing the charge-$3$ scalar field \cite{fradkin-shenker-1979}. Hence, we show that there is no Chern-Simons terms for $\alpha_\mu$ and $A_\mu$ on the surface. 

We start with the theory in which the charge-$3$ scalar field does not condense but the partons $\psi_j$ form the topological band insulator and the U(1) gauge field $\alpha_\mu$ is in the deconfined Coulomb phase. Then the bulk action for this topological phase is 
\begin{align}
\mathcal{L} = \frac{3}{32\pi} \epsilon^{\mu\nu\lambda\rho} (\frac{1}{3}F_{\mu\nu} + f_{\mu\nu}) (\frac{1}{3}F_{\lambda\rho} + f_{\lambda\rho}), 
\end{align}
in which $f_{\mu\nu} = \partial_\mu \alpha_\nu - \partial_\nu \alpha_\mu$ and $F_{\mu\nu}$ is the field strength of $A_\mu$. On the surface, this bulk action contributes the Chern-Simons term
\begin{align}
\mathcal{L} = \frac{3}{8\pi} \epsilon^{\mu\nu\lambda} (\frac{1}{3}A_\mu + \alpha_\mu) \partial_\nu (\frac{1}{3}A_\lambda + \alpha_\lambda).  
\end{align}
This contribution, however, is exactly cancelled by the regularization contribution of the Dirac fermions living on the surface of this topological phase. To see explicitly, we note that the regularized theory of the surface Dirac fermions is  
\begin{align}
\mathcal{L} = \sum_{i=1}^{3} i\bar{\psi}_j \slashed{D}_{\frac{1}{3}A + \alpha} \psi_j - \frac{3}{8\pi} \epsilon^{\mu\nu\lambda} (\frac{1}{3}A_\mu + \alpha_\mu) \partial_\nu (\frac{1}{3}A_\lambda + \alpha_\lambda), 
\end{align}
where the second Chern-Simons term is the contribution from the regularization term, which is exactly the opposite of the bulk contribution. Hence, on the surface of this topological phase with the bulk contribution and regularization contribution, we end up with the following theory 
\begin{align}
\mathcal{L} = \sum_{i=1}^{3} i\bar{\psi}_j \slashed{D}_{\frac{1}{3}A + \alpha} \psi_j.  
\end{align}
Now, we condense the charge-$3$ scalar field, which will break the U(1) gauge symmetry of the field $\alpha_\mu$ to its discrete subgroup $\mathbb{Z}_3$. This condensation does not break  time-reversal invariance nor changes the topological band structure of the fermionic partons. Thus, we do not expect to have any topological terms to appear on the surface after  condensation. Because the condensation will gap out the fluctuations of the gauge field $\alpha_\mu$, we can ignore its fluctuations at the lowest energies, and find 
\begin{align}
\mathcal{L} = \sum_{i=1}^{3} i\bar{\psi}_j \slashed{D}_{\frac{1}{3}A} \psi_j,   
\end{align}
which is the Eq. (2) in the main text. Note that there is no Chern-Simons term for the $A$ and the $\mathbb{Z}_3$ gauge fields. 

\end{document}